\documentclass[%
 reprint,
 amsmath,amssymb,
 aps,
pra,
]{revtex4-2}

\usepackage{graphicx}
\usepackage{dcolumn}
\usepackage{bm}

\newcommand{\lH}{\mathcal{H}}
\newcommand{\lD}{\mathcal{D}}

\usepackage{physics}

\usepackage{xcolor,tikz,pgfplots}
\usepackage[siunitx]{circuitikz}
\usetikzlibrary{circuits,arrows,calc,intersections,decorations,decorations.pathreplacing, decorations.pathmorphing,lindenmayersystems,patterns}	

\usepackage{dsfont}
\usepackage{float}

\begin{document}

\title{A microscopic revision of the rate equations of a heavily erbium-doped fiber laser}

\author{Iaroslav L. Pozdniak$^{1,3,4}$}
 \email{pozdniak.ial@phystech.edu}
\author{Ivan V. Vovchenko$^{1,2}$}
\author{Evgeny S. Andrianov$^{1,2,3}$}
\author{Alexander V. Dorofeenko $^{1,2,3}$}
\affiliation{%
${}^1$Moscow Institute of Physics and Technology, 141700, 9 Institutskiy pereulok, Moscow, Russia
}%
\affiliation{%
${}^2$Dukhov Research Institute of Automatics, 127055, 22 Sushchevskaya, Moscow, Russia
}%
\affiliation{%
${}^3$Institute for Theoretical and Applied Electromagnetics, 125412, 13 Izhorskaya, Moscow, Russia
}%
\affiliation{%
${}^4$Kotelnikov Institute of Radioengineering and Electronics of RAS, 125009, 11 Mokhovaya, Moscow, Russia
}%

\date{\today}

\begin{abstract}
Heavily erbium-doped fiber lasers can exhibit self-pulsing induced by interacting ion pairs, which was found in numerous experiments and supported by modeling. However, conventional rate-equation models rely on phenomenological parameters with an unclear microscopic origin. We derive a rate-equation model using a Lindblad master equation describing isolated erbium ions and interacting ion pairs coupled to a single cavity mode, with cooperative upconversion represented by an effective dissipative process. In the strong-dephasing limit, adiabatic elimination of coherences and ensemble averaging yield a closed system of equations for the cavity photon number, isolated-ion inversion, and pair populations. We show that in the same limit the pair–cavity coupling constant reduces to that of two independent ions, without superradiant or subradiant contributions. The resulting equations retain spontaneous emission, which provides a photon-number floor between pulses and substantially affects the laser pulsing dynamics. Numerical analysis reveals self-pulsing within finite pump-rate intervals and identifies the region of self-pulsing as a function of the upconversion rate and pair fraction. The model thus establishes a microscopic basis for analyzing self-pulsing in heavily erbium-doped fiber lasers.
\end{abstract}

\maketitle

\section{\label{sec:level1} Introduction}

Erbium-doped fibers (EDFs) are widely used in the telecommunication spectral range around $1.5~\mu\text{m}$ due to their efficient gain and lasing based on the ${}^4I_{13/2}\rightarrow{}^4I_{15/2}$ transition, as well as compatibility with standard silica fibers. When the concentration of $\mathrm{Er}^{3+}$ ions is relatively low, the erbium-doped fiber laser (EDFL) can be effectively described as an ensemble of effective two-level systems (TLS) with strong dephasing. This leads to the class-B laser rate equations for the population inversion and the number of photons in the cavity \cite{Lacot1994}.

However, at higher concentrations erbium ions tend to form pairs and clusters with their own internal dynamics that affect the optical properties of the medium. To take this into account, two phenomenological rate-equation approaches were suggested: \textit{pair-induced quenching} (PIQ) and \textit{cooperative upconversion} (CUC) \cite{Delevaque1993,Snoeks1995}. They share the same microscopic motivation: when two paired erbium ions are excited, energy transfer ${}^4I_{13/2}+{}^4I_{13/2}\rightarrow{}^4I_{15/2}+{}^4I_{9/2}$ can occur, which, from the perspective of the effective TLS, establishes an additional nonlinear relaxation channel.

A systematic investigation of the connection between PIQ and the self-pulsing behavior of EDFLs was made in the classical papers of Le Boudec and Sanchez \cite{LeBoudec1993,Sanchez1993}. They reported that, depending on the parameters, an EDFL can operate in steady-state, sinusoidal, or self-pulsing regimes. A model of the gain medium as a mixture of isolated ions and pairs reproduced this behavior: within a certain pump range the steady state loses stability and a limit cycle appears, with the boundaries of the range corresponding to Hopf bifurcations. Further investigations showed that the same rate-equation model also reproduces antiphase dynamics and chaotic behavior \cite{Sanchez1996}.

Despite the success of these rate-equation approaches, their connection with the microscopic physics remains unclear, and some phenomenological coefficients lack a clear interpretation and have to be fitted to every particular experiment -- a notable example being the pair cross-section ratio $y=\sigma'_\ell/\sigma_\ell\simeq 0.2$ required in the Sanchez model \cite{Sanchez1993}. Moreover, density-matrix treatments of cooperative upconversion have revealed pair-dynamics effects that are in principle invisible at the rate-equation level --- population pulsations and instabilities that may explain the observed loss of stability of the
2.8-$\mu$m pair-pumped Er laser \cite{Nonlinear1994}. Those effects arise in the regime where the interatomic coupling exceeds the dephasing rate; in heavily doped fibers at room temperature the hierarchy is reversed. Nevertheless, the validity and the structure of the rate equations themselves must be established by a consistent microscopic derivation rather than postulated.

A consistent microscopic derivation of the EDFL rate equations for isolated ions was recently proposed by Vovchenko et al. \cite{Vovchenko2024}, where a master-equation description of the mode-ion ensemble produced, in particular, microscopic spontaneous-emission terms smoothing out the lasing threshold. Subsequently, a full-quantum description including pairs of ions was developed in \cite{Vovchenko2025}; there the pair dynamics originates from the coherent exchange interaction and the detuning between the ions, while cooperative upconversion is not included and no reduction to rate equations is performed.

In this paper we derive the rate equations of a heavily erbium-doped fiber laser microscopically, with cooperative upconversion of ion pairs as the central process involving pairs. Starting from a master equation for a pair coupled to the cavity mode, we propose an effective dissipative upconversion channel and distinguish the symmetric and antisymmetric coupling channels of the pair to the field. Then, after adiabatic elimination and ensemble averaging, we obtain a closed set of rate equations in which the contributions of isolated ions and pairs enter on equal microscopic footing. The resulting equations reproduce the self-pulsing dynamics and provide a microscopic phase diagram of the instability in the pair-parameters plane.

\section{Microscopic approach to modeling the EDFL}

In this paper, we develop a model of a heavily erbium-doped single-mode fiber laser. The single lasing mode can be realized in Bragg fiber grating structures that form Fabry-Perot or distributed-feedback cavities \cite{Yamashita1998,Babin2007}. At high erbium concentrations, ions may form clusters, affecting the EDFL dynamics. Following \cite{Sanchez1993}, we independently describe (i) isolated ions as effective TLSs and (ii) interacting ion pairs exhibiting cooperative upconversion. For both groups of ions, we propose a quantum model and use the Lindblad master-equation formalism to obtain microscopically consistent rate equations. In particular, it allows us to correctly take into account the interaction of the ions with the cavity mode, including spontaneous emission, which plays a crucial role in the self-pulsing regime.

Throughout the paper, we follow the same derivation pipeline for isolated ions and ion pairs, which we refer to as "building blocks":
\begin{enumerate}
\item Microscopic quantum model of a single building block of the gain medium: the Hamiltonian, including the cavity mode and the light--matter interaction, and the jump operators of the dissipative processes.
\item Dynamical equations for a chosen set of observables (mean values of operators), closed by mean-field factorization of higher-order correlators, with adiabatic elimination of fast (slave) variables based on the rate hierarchy.
\item Ensemble averaging over the building blocks of the same type.
\item Combining several types of building blocks (isolated ions and pairs) into a single coupled system of rate equations.
\end{enumerate}

The Lindblad master equation \cite{breuer2002theory} for the density operator $\rho$ reads
\[ \dot{\rho} = -\frac{i}{\hbar} [\lH, \rho] + \sum_k \lD_{L_k}[\rho],\]
where $\lH$ is the Hamiltonian of the system and $L_k$ are the jump operators of processes corresponding to non-unitary (dissipative) dynamics, and $\lD_{L_k}$ is a Lindblad dissipator
\[\lD_{L_k}[\rho]
=
L_k \rho L_k^\dagger - \frac{1}{2} \left\lbrace L_k^\dagger L_k, \rho \right\rbrace.\]
It implies the dynamical equation for the mean value of any operator $O$:
\begin{equation}\label{meanvaluedynamics}
\dot{\langle O \rangle} = \frac{i}{\hbar} \langle [\lH, O] \rangle
+ \sum_k \langle \lD^\dagger_{L_k}[O] \rangle ,\end{equation}
with
\[
\lD^\dagger_{L_k}[O] = -\frac{1}{2} \left( L_k^\dagger [L_k,O] + [O,L_k^\dagger]L_k \right).
\]
Simultaneous commutation $[L_k,O]=[L_k^\dagger,O]=0$ is a sufficient condition for $\lD^\dagger_{L_k}[O]=0$. We use this property extensively: for an observable of a given building block all dissipative terms of the other blocks vanish, so the dynamical equations of independent blocks do not affect each other and can be averaged over the ensemble directly.

\section{Rate equations for an ensemble of isolated ions}

In this section, we apply the approach outlined above to derive the standard rate equations for a class-B laser. This well-established example serves to introduce the notation, illustrate the approximations involved in the derivation, and provide a basis for the analysis of a more complex laser system in the following section.

We consider a single-mode cavity of frequency $\omega_c + i \gamma/2$. The cavity mode is quantized according to the standard procedure, with operators $a, a^\dagger$ forming the bosonic algebra $[a, a^\dagger] = 1$.

The cavity wavelength $\lambda_c \approx 1.55~\mu\text{m}$ is close to the wavelength of the $^4I_{13/2} \to {}^4I_{15/2}$ transition. A laser with a wavelength $\lambda_p \approx 0.98~\mu\text{m}$ acts as the pump. It excites the $^4I_{15/2} \to {}^4I_{11/2}$ transition, after which the state ${}^4I_{11/2}$ rapidly relaxes to ${}^4I_{13/2}$. Thus, $\rm Er^{3+}$ is modeled as an effective TLS pumped incoherently.

Let us call the eigenstates of the TLS $\ket{e}$ and $\ket{g}$, and the transition energy $\hbar \omega_0$. The operators $\sigma_{\pm,z}$ form the standard spin-1/2 algebra: $[\sigma_+, \sigma_-] = \sigma_z$, $[\sigma_z, \sigma_{\pm}] = \pm 2 \sigma_{\pm}$. A Jaynes–Cummings Hamiltonian appears:

\[ \lH =
\underbrace{\hbar \omega_c a^\dagger a}_\text{cavity} +
\underbrace{\hbar \omega_0 \sigma_{+} \sigma_{-}}_\text{ion} +
\underbrace{\hbar \Omega_l (\sigma_{+} a + a^\dagger \sigma_{-})}_\text{interaction}, \]
where $\Omega_l$ is the \textit{local} Rabi frequency. It depends on the dipole moment $d_{eg}$ of the transition $\ket{e} \leftrightarrow \ket{g}$, the amplitude $E$ of the electric field of a single photon, and the angle $\theta$ between the dipole moment and the electric field via $\hbar \Omega_l = \hbar \Omega \cos \theta = d_{eg} E \cos \theta$. This angle $\theta$ is fixed for every ion due to the Stark splitting in the randomly oriented crystal field of glass. 

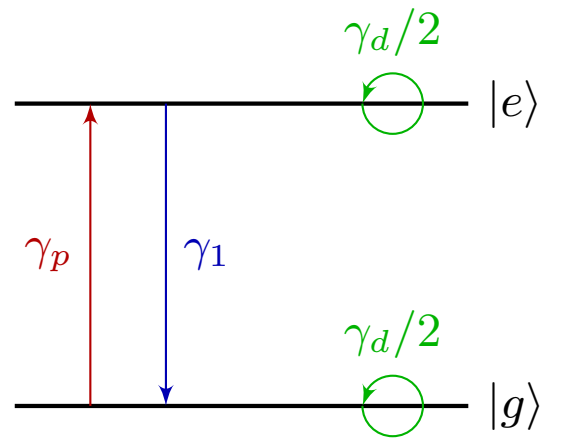
\begin{figure}[t]
\begin{tikzpicture}[>=latex']
	\draw [thick] (0,0) -- (3,0) node [right] {$\ket{g}$}; 
	\draw [thick] (0,2) -- (3,2) node [right] {$\ket{e}$}; 
	
	\draw[red!70!black,->] (.5,0) -- (.5,2) node [pos=.5,left] {$\gamma_p$};
	\draw[blue!70!black,->] (1.0,2) -- (1.0,0) node [pos=.5,right] {$\gamma_1$};

	\draw[green!70!black,->] (2.3,0) arc (-180:180:.2);
	\path (2.5,0) node [green!70!black,above=.2cm] {$\gamma_d/2$};
	\draw[green!70!black,->] (2.3,2) arc (-180:180:.2);
	\path (2.5,2) node [green!70!black,above=.2cm] {$\gamma_d/2$};
\end{tikzpicture}
\caption{\label{fig:isolated-ion-processes}Schematic representation of the non-unitary processes for an isolated ion.}
\end{figure}

Four processes are sufficient \cite{Vovchenko2024} to describe this system's non-unitary dynamics (Fig.~\ref{fig:isolated-ion-processes}):
\begin{enumerate}
\item Cavity mode loss with a jump operator  $\sqrt{\gamma} \, a$
\item Incoherent pump with a jump operator $ \sqrt{\gamma_p} \, \sigma_{+}$
\item Relaxation of the excited state (nonradiative and radiative to other optical modes) with a jump operator $\sqrt{\gamma_1} \, \sigma_{-}$
\item TLS dephasing with a jump operator $\sqrt{\frac{\gamma_d}{2}} \,  \sigma_{z}$,
\end{enumerate}
where $\gamma$, $\gamma_p$, $\gamma_1$, and $\gamma_d$ are the corresponding rates. 

\subsection{Dynamical equations and adiabatic elimination}
A minimal representative set of dynamical variables that yields a closed system is: the number $n_p = \langle a^\dagger a \rangle$ of photons in the cavity; the population inversion $D = \langle \sigma_z \rangle$; and the correlator $J = \langle \sigma_+ a \rangle$ of the ion and the cavity field \cite{Scully}. After direct substitution into \eqref{meanvaluedynamics} and truncation (mean-value factorization) $\langle a^\dagger a\, \sigma_z \rangle \simeq \langle a^\dagger a \rangle \langle \sigma_z \rangle$ we obtain the system:
\begin{equation}\label{original-single-ion}
\begin{cases}
\dot{n}_p = -\gamma n_p - 2 \Omega_l \Im J \\
\dot{D} = -\gamma_1 (1+D) + \gamma_p (1-D) + 4 \Omega_l \Im J \\
\dot{J} = - (\Gamma + i \, \delta \omega) J - i \Omega_l \left( n_pD + \frac{1+D}{2} \right)
\end{cases},
\end{equation}
where $\Gamma = \dfrac{\gamma + \gamma_p + \gamma_1}{2} + \gamma_d$, $\delta \omega = \omega_c - \omega_0$.

For the adiabatic elimination we have to establish a hierarchy of rates. The cavity loss rate $\gamma$ is determined by the cavity $Q$--factor and internal losses; typically $\gamma \approx 10^8~\text{s}^{-1}$. The relaxation of the excited state can be determined from photoluminescence decay curves \cite{Snoeks1995}; typically $\gamma_1 = 10^{2}~\text{s}^{-1}$. The dephasing rate can be determined from fluorescence line narrowing or spectral hole burning; typically $\gamma_d = 10^{12}~\text{s}^{-1}$ at room temperature. These rates form the hierarchy $\gamma_d \gg \gamma \gg \gamma_{1}, \gamma_p$.
    
Thus, the relaxation rate $\Gamma \approx \gamma_d$ of $J$ in \eqref{original-single-ion} is much larger than others. This allows us to treat $J$ as a slave variable (adiabatic elimination $\dot{J}=0$) and obtain:

\[ 2\Omega_l \Im J = -\underbrace{\frac{2\Gamma \Omega_l^2}{\Gamma^2 + \delta \omega^2}}_{\xi_l}  \left(n_pD + \frac{1+D}{2}\right),\]
where $\xi_l$  is the \textit{local} coupling constant of a single ion with the cavity. Finally, for a single ion we obtain the rate equations:
\begin{equation}\label{final-single-ion}
\begin{cases}
\dot{n}_p = -\gamma n_p + \xi_l \left(n_pD + \frac{1+D}{2}\right)  \\
\dot{D} = -\gamma_1 (1+D) + \gamma_p (1-D) - 2 \xi_l \left(n_pD + \frac{1+D}{2}\right)
\end{cases},
\end{equation}

\subsection{Ensemble averaging}
Let $N$ be the total number of ions in the cavity. Dynamical equations \eqref{original-single-ion} for different ions are mostly independent due to the cross-commutation $[\sigma_{ai},\sigma_{bj}]=0$, $a,b \in \{\pm,z\}$. The only "interaction" term appears when we calculate $\langle \dot{J} \rangle$ according to \eqref{meanvaluedynamics} in commutators $\langle [\lH, \sigma_+ a ] \rangle$. It is proportional to $\hbar \Omega\langle \sigma_{j-} \sigma_{k+} \rangle$ and describes energy exchange between ions through the cavity mode. We neglect these terms due to the weakness of the ion-cavity interaction and strong dephasing. Thus, the equation for $D$  in \eqref{final-single-ion} also applies to the ensemble average $\sum D_i / N$ if we use the approximation $\langle \xi_l D \rangle \simeq  \langle \xi_l \rangle \langle D \rangle$. At the same time, every atom interacts with the cavity independently and contributes to $\langle [\lH, a^\dagger a] \rangle$ when we write a dynamical equation according to \eqref{meanvaluedynamics}. Due to the random direction of the dipole moment, the value of $\xi_l$ is averaged over the ensemble: 
\[ \xi = \langle \xi_l \rangle = \left\langle \frac{2 \Gamma \Omega^2}{\Gamma^2 + \delta \omega^2} \cos^2 \theta \right\rangle = \frac{2\Gamma \Omega^2}{\Gamma^2 + \delta \omega^2}  \cdot \frac{1}{3}.\]

After summation over the ions and ensemble averaging, the dynamical system reads
\begin{equation}\label{final-single-ions}
\begin{cases}
\dot{n}_p = -\gamma n_p +  N \xi \left(n_pD + \frac{1+D}{2}\right)  \\
\dot{D} = -(\gamma_1 + \xi) (1+D) + \gamma_p (1-D) - 2 \xi n_pD
\end{cases}
\end{equation}
and it coincides with the class-B laser rate equations \cite{Lacot1994,Siegman1986}. For every value of the pump rate $\gamma_p$ it has only one stationary point $(n_p^*,D^*)$, which is always stable. For $\gamma_p \gg \gamma_1$ the angular frequency of relaxation oscillations around it is given by $\sqrt{\gamma_1 N \xi\left(\frac{\gamma_p}{\gamma_1}-1\right)}$ for $\gamma \ll N \xi$. We note that the quantum treatment of the cavity yields the cavity--medium interaction term in the form $N \xi \left(n_p D + \frac{1+D}{2} \right)$, which contains both stimulated and spontaneous emission. The spontaneous emission correction in the equation for $D$ can be absorbed into a redefinition of $\gamma_1$, but it plays an important role in the first equation as the floor from which the photon number grows.

\section{Rate equations for an ensemble of ion pairs}
In this section we use the proposed approach to obtain the rate equations for an ensemble of ion pairs. We assume that a pair consists of two identical ions located close enough that the dipole-dipole interaction affects their individual behavior. 

To take cooperative upconversion into account, we consider three levels of each erbium atom: $\ket{g} = {}^4I_{15/2}$, $\ket{e} = {}^4I_{13/2}$, and $\ket{u} = {}^4I_{9/2}$. It is well known \cite{Georges1992,Sanchez1993} that the total energies of two erbium ions in the states ${}^4I_{9/2} + {}^4I_{15/2}$ and ${}^4I_{13/2} + {}^4I_{13/2}$ are close. Consequently, the dipole-dipole interaction causes a phonon-assisted transition $\ket{ee} \leftrightarrow \ket{gu/ug}$. Then, the state $\ket{u}$ rapidly relaxes to $\ket{e}$ through a multiphonon channel \cite{Layne1977}. This allows us to retain the TLS approximation for each ion and describe the resulting process as a non-unitary transition $\ket{ee} \to \ket{eg/ge}$ with rate $\gamma_u$.

A Jaynes–Cummings Hamiltonian modified by the Förster exchange term for a pair of identical ions reads
\begin{equation}\label{pairHamiltonian}
\begin{split}
\lH =
\underbrace{\hbar \omega_c a^\dagger a}_\text{cavity} 
+
\underbrace{\sum_j^2 \hbar \Omega \cos \theta_j \left( \sigma_{+j} a + \sigma_{-j} a^\dagger \right)}_\text{interaction}
+ \\  +
\underbrace{\sum_j^2 \hbar \omega_0 \sigma_{+j} \sigma_{-j}
+
\hbar J_{dd} \left( \sigma_{+1}\sigma_{-2} + \sigma_{-1}\sigma_{+2} \right)}_\text{pair},
\end{split}
\end{equation}
where $j$ is the number of the ion in the pair. We emphasize that the ions are affected by different crystal electric fields, so their dipole moments make different angles $\theta_j$ with the electric field of the cavity mode.

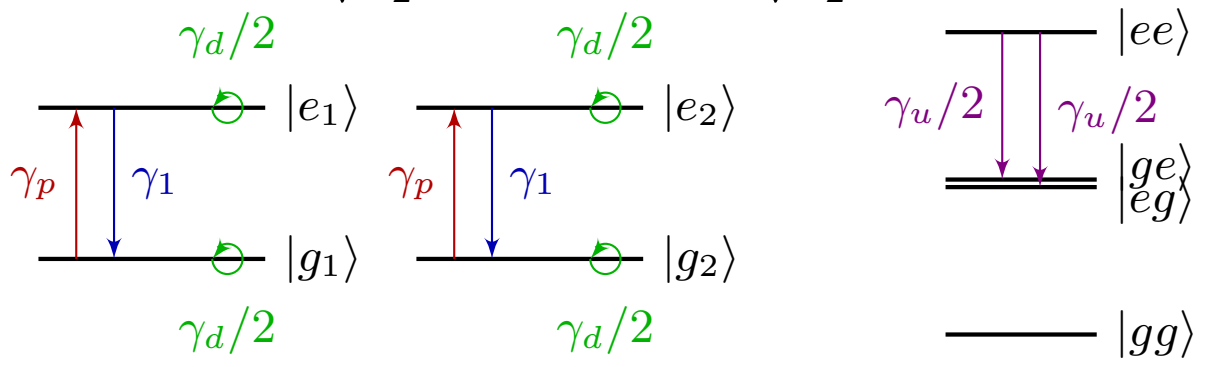
\begin{figure}[H]
\centering
\begin{tikzpicture}[>=latex',scale=0.5]

	\draw [thick] (0,0) -- (3,0) node [right] {$\ket{g_1}$}; 
	\draw [thick] (0,2) -- (3,2) node [right] {$\ket{e_1}$}; 
	
	\draw[red!70!black,->] (.5,0) -- (.5,2) node [pos=.5,left] {$\gamma_p$};
	\draw[blue!70!black,->] (1,2) -- (1,0) node [pos=.5,right] {$\gamma_1$};

	\draw[green!70!black,->] (2.3,0) arc (-180:180:.2);
	\path (2.5,0) node [green!70!black,below=.2cm] {$\gamma_d/2$};
	\draw[green!70!black,->] (2.3,2) arc (-180:180:.2);
	\path (2.5,2) node [green!70!black,above=.2cm] {$\gamma_d/2$};

	\begin{scope}[shift={(5,0)}]
	\draw [thick] (0,0) -- (3,0) node [right] {$\ket{g_2}$}; 
	\draw [thick] (0,2) -- (3,2) node [right] {$\ket{e_2}$}; 
	
	\draw[red!70!black,->] (.5,0) -- (.5,2) node [pos=.5,left] {$\gamma_p$};
	\draw[blue!70!black,->] (1,2) -- (1,0) node [pos=.5,right] {$\gamma_1$};

	\draw[green!70!black,->] (2.3,0) arc (-180:180:.2);
	\path (2.5,0) node [green!70!black,below=.2cm] {$\gamma_d/2$};
	\draw[green!70!black,->] (2.3,2) arc (-180:180:.2);
	\path (2.5,2) node [green!70!black,above=.2cm] {$\gamma_d/2$};
	\end{scope}

	\begin{scope}[shift={(12,-1)}]
	\draw [thick] (0,0) -- (2,0) node [right] {$\ket{gg}$}; 
	\draw [thick] (0,2.05) -- (2,2.05) ++ (0,.17) node [right] {$\ket{ge}$}; 
	\draw [thick] (0,1.95) -- (2,1.95) ++ (0,-.17) node [right] {$\ket{eg}$}; 
	\draw [thick] (0,4) -- (2,4) node [right] {$\ket{ee}$}; 
	
	\draw[red!50!blue,->] (.75,4) -- (.75,2.05) node [pos=.5,left] {$\gamma_u/2$};
	\draw[red!50!blue,->] (1.25,4) -- (1.25,1.95) node [pos=.5,right] {$\gamma_u/2$};
	\end{scope}

\end{tikzpicture}
\caption{\label{fig:ion-pair-processes}Schematic representation of the non-unitary processes for an ion pair.}
\end{figure}

From the perspective of non-unitary dynamics, we keep everything that is sufficient for individual ions and add cooperative upconversion (Fig.~\ref{fig:ion-pair-processes}):

\begin{enumerate}
\item Cavity mode loss with a jump operator $\sqrt{\gamma} \, a$
\item Two individual incoherent pumps with jump operators $\sqrt{\gamma_p} \, \mathds{1}_{1} \otimes \sigma_{+2}$ and $\sqrt{\gamma_p} \, \sigma_{+1} \otimes \mathds{1}_{2}$
\item Two individual relaxations of the excited state with jump operators $\sqrt{\gamma_1} \, \mathds{1}_{1} \otimes \sigma_{-2}$ and $\sqrt{\gamma_1} \, \sigma_{-1} \otimes \mathds{1}_{2}$
\item Two individual TLS dephasing with jump operators $\sqrt{\frac{\gamma_d}{2}} \, \mathds{1}_{1} \otimes \sigma_{z2}$ and $\sqrt{\frac{\gamma_d}{2}} \, \sigma_{z1} \otimes \mathds{1}_{2}$.
\item Two cooperative upconversion channels with jump operators $\sqrt{\frac{\gamma_u}{2}} \, \ket{eg}\bra{ee}$ and $\sqrt{\frac{\gamma_u}{2}} \, \ket{ge}\bra{ee}$
\end{enumerate}

For a system of identical spins it is natural to use the basis $\ket{G} = \ket{gg}$, $\ket{E} = \ket{ee}$, $\ket{\pm} = (\ket{eg} \pm \ket{ge})/\sqrt{2}$. We use the notation $\sigma_{AB} = \ket{A}\bra{B}$ for the transition operators, and $P_A = \ket{A}\bra{A}$ for the projectors, where $A,B\in\{G,E,+,-\}$. In this notation the Hamiltonian \eqref{pairHamiltonian} reads
\[
\begin{split}
\lH  = \hbar \omega_c a^\dagger a
+
\hbar \omega_0 \left( P_+ + P_- + 2P_E \right)
+
\hbar J_{dd} (P_+ - P_-)
+ \\
+
\hbar \Omega_s \left( \sigma_{+G} + \sigma_{E+} \right) a
+
\hbar \Omega_a \left( \sigma_{-G} - \sigma_{E-} \right) a + \text{h.c.}
\end{split}
.\]
Here $\Omega_s = \Omega (\cos \theta_1 + \cos \theta_2)/\sqrt{2}$ is the Rabi frequency of the symmetric channel, and $\Omega_a = \Omega (\cos \theta_1 - \cos \theta_2)/\sqrt{2}$ is that of the antisymmetric channel.

\subsection{Dynamical equations}
It is natural to choose the following variables: the number of photons $n_p = \langle a^\dagger a \rangle$ in the cavity; the populations $N_A = \langle P_A \rangle$; the correlators $J_{AB} = \langle \sigma_{AB} a \rangle$ of ions and the cavity field, $A$ and $B$ denoting $E$, $G$, $+$ and $-$. The dynamical equations for the number of photons $n_p$ and populations $N_A$ do not require an extension of the proposed set of dynamical variables:
\begin{widetext}
\begin{equation}\label{np-NA-dynamics}
\begin{split}
\dot{n}_p &= - \gamma n_p - 2 \Omega_s  \left( \Im J_{+G} + \Im J_{E+} \right) - 2 \Omega_a \left( \Im J_{-G} - \Im J_{E-} \right)
\\
\dot{N}_E &= 2 \Omega_s \Im J_{E+} - 2 \Omega_a \Im J_{E-} + \gamma_p (N_+ + N_-) - 2\gamma_1 N_E - \gamma_u N_E
\\
\dot{N}_+ &= 2 \Omega_s \Im J_{+G} - 2 \Omega_s \Im J_{E+}	 + \gamma_p (N_G - N_+) + \gamma_1 (N_E - N_+) + \gamma_d ( N_- - N_+) + \frac{\gamma_u}{2} N_E
\\
\dot{N}_- &= 2 \Omega_a \Im J_{-G} + 2 \Omega_a \Im J_{E-} + \gamma_p (N_G - N_-) + \gamma_1 (N_E - N_-) - \gamma_d (N_- - N_+) + \frac{\gamma_u}{2} N_E
\\
\dot{N}_G &= -2 \Omega_s \Im J_{+G} - 2 \Omega_a \Im J_{-G} - 2 \gamma_p N_G + \gamma_1 (N_+ + N_-)
\end{split},
\end{equation}
\end{widetext}

which obviously satisfy the normalization condition $N_E + N_+ + N_- + N_G = 1$. The dynamical equations for the correlators $J_{AB}$ read

\begin{widetext}
\begin{equation}\label{pair-correlators-dynamics}
\begin{split}
\dot{J}_{E+} &= -\left(\Gamma_{E\pm} + i \delta \omega_+ \right) J_{E+} + i \Omega_s \left( \langle P_+ a^\dagger a \rangle - \langle P_E a a^\dagger \rangle - \langle \sigma_{EG} a^2 \rangle  \right)- i \Omega_a \langle \sigma_{-+} a^\dagger a \rangle + \gamma_p J_{+G}
\\
\dot{J}_{E-} &= -\left(\Gamma_{E\pm} + i \delta \omega_- \right) J_{E-} + i \Omega_s \langle \sigma_{+-} a^\dagger a \rangle + i \Omega_a \left( \langle P_E a a^\dagger \rangle - \langle P_- a^\dagger a \rangle - \langle \sigma_{EG} a^2 \rangle \right) - \gamma_p J_{-G}
\\
\dot{J}_{+G} &=-\left(\Gamma_{\pm G} + i \delta \omega_- \right) J_{+G} + i \Omega_s \left( \langle P_G a^\dagger a \rangle - \langle P_+ a a^\dagger \rangle + \langle \sigma_{EG} a^2 \rangle \right) - i \Omega_a \langle \sigma_{+-} a a^\dagger \rangle  + \gamma_1 J_{E+}
\\
\dot{J}_{-G} &= -\left(\Gamma_{\pm G} + i \delta \omega_+ \right) J_{-G} - i \Omega_s \langle \sigma_{-+} a a^\dagger \rangle + i \Omega_a \left( \langle P_G a^\dagger a \rangle - \langle P_- a a^\dagger \rangle - \langle \sigma_{EG} a^2 \rangle \right)  - \gamma_1 J_{E-}
\end{split}
,\end{equation}
\end{widetext}

where $\Gamma_{E\pm} = \frac{\gamma}{2} + \frac{\gamma_p}{2} + \frac{3\gamma_1}{2} + \gamma_d + \frac{\gamma_u}{2}$ , $\Gamma_{\pm G} = \frac{\gamma}{2} + \frac{3 \gamma_p}{2} + \frac{\gamma_1}{2} + \gamma_d$, $\delta \omega_{\pm} = \delta \omega \pm J_{dd}$. Most of the terms that appear here can be factorized, $\langle P_A a^\dagger a \rangle \simeq N_A n_p$. However, two additional variables should be discussed: the correlators of the symmetric and antisymmetric states $C_{\pm} = \langle \sigma_{+-} \rangle$ and of the two-photon interaction $J_{EG} = \langle \sigma_{EG} a^2 \rangle$. Their dynamics is given by

\begin{widetext}
\[
\begin{split}
&\dot{C}_{\pm} = 2i J_{dd}C_{\pm} + i \Omega_s \left( J_{E-} + J^*_{-G} \right) + i \Omega_a \left( - J_{+G} + J^*_{E+} \right) - (\gamma_p+\gamma_1) C_{\pm} - \gamma_d (C_{\pm} - C_{\pm}^*)
\\
&\dot{J}_{EG} = - (\Gamma_{EG} + 2i \delta \omega) J_{EG} + i \Omega_s (\langle \sigma_{+G}a^\dagger a^2 \rangle -\langle \sigma_{E+} a^2 a^\dagger \rangle ) + i\Omega_a (\langle \sigma_{-G} a^\dagger a^2 \rangle - \langle \sigma_{E-} a^2 a^\dagger \rangle )
\end{split},
\]
\end{widetext}

where $\Gamma_{EG} = \gamma + \gamma_p + \gamma_1 + 2 \gamma_d + \frac{\gamma_u}{2}$. After the factorization, we have a closed dynamical system for 11 variables: $n_p$; the four $N_A$; the five $J_{AB}$, and $C_\pm$.

\subsection{Adiabatic elimination}
All dynamical equations for the correlators $J_{AB}$ have a relaxation term $\dot{J}_{AB} = - \Gamma_{AB} J_{AB}$ with $\Gamma_{AB} \approx \gamma_d$. Consequently, all $J_{AB}$ can be declared slave variables and adiabatically eliminated with $\dot{J}_{AB}=0$. The same relaxation is observed for $\Im C_\pm$, so we assume that $C_\pm = C^*_\pm$.

In the equations \eqref{pair-correlators-dynamics} for $J_{AB}$ the term $J_{EG}$ competes with the terms $N_A n_p$. Using the adiabatic elimination of $J_{EG}$
\[
\begin{split}
J_{EG} \approx  n_p\frac{i \Omega_s ( J_{+G} - J_{E+} ) - i\Omega_a ( J_{-G} + J_{E-})}{(\Gamma_{EG} + 2i \delta \omega) }
\end{split}
\]

we can estimate that  $|J_{EG}| \approx n_p J_{AB} \cdot \Omega/\Gamma \approx \Omega^2 n_p/\Gamma^2 \cdot N_A n_p \approx \xi n_p/\Gamma \cdot N_A n_p$. For an order-of-magnitude estimate, we take $\xi\approx10^{-4}~\mathrm{s}^{-1}$ (see Table~\ref{tab:parameters});  $\Gamma \approx \gamma_d \approx 10^{12}~\text{s}^{-1}$ at room temperature; for a typical EDFL emitting several milliwatts one can estimate $n_p \approx 10^{9}$. Substitution shows that under these conditions $J_{EG} \approx 10^{-7} N_A  n_p$, so $J_{EG}$ is a small correction and can simply be neglected.

We also note that $N_- - N_+$ relaxes with a rate of $2 \gamma_d$,  which greatly
exceeds the rates of the pump, population relaxation, and field-driven
population dynamics. We therefore consider that $N_+ \simeq N_- = N_S$. Under this assumption, the normalization condition reads $N_E + 2 N_S + N_G = 1$.

The next step is to solve the algebraic equations for the correlators $J_{AB}$. First, we neglect the terms $\gamma_{1} J_{AB}$, $\gamma_p J_{AB}$ because they contribute to the rate equations only as very small terms of order $\gamma_{1,p} \Omega^2/\Gamma^2$. Second, we use the approximations $\Gamma_{E \pm} \simeq \Gamma_{\pm G} \simeq \Gamma \simeq \gamma_d$ and $\delta \omega_\pm \simeq \delta \omega$. Third, we introduce $g_{a,s} = \Omega_{a,s} \Gamma/(\Gamma^2 + \delta \omega^2)$. The resulting equations read:
\[
\begin{split}
\Im J_{E+}
&=
g_s \left( n_pN_S - (n_p+1)N_E   \right)-  g_a  n_p C_\pm
\\
\Im J_{E-}
&=
g_a \left( (n_p+1)N_E - n_pN_S  \right) + g_s n_p C_{\pm}
\\
\Im J_{+G}
&=
g_s \left( n_pN_G - (n_p+1)N_S \right) - g_a (n_p+1) C_{\pm}
\\
\Im J_{-G}
&= 
g_a \left( n_pN_G - (n_p+1)N_S  \right)  - g_s (n_p+1) C_\pm
\end{split}.
\]
Since $g_a\Omega_s=g_s \Omega_a$, we substitute $J_{AB}$ into \eqref{np-NA-dynamics} and obtain a system of five rate equations for a single pair, i.e. for the variables $n_p$, $N_{E,G,S}$ and $C_\pm$.  The equation for $C_{\pm}$ is eliminated later, after ensemble averaging, but the first four read:
\begin{widetext}
\begin{equation}\label{final-single-pair}
\begin{cases}
\dot{n}_p = -\gamma n_p + 2 (\Omega_s g_s +\Omega_a g_a) \left[ (n_p+1)N_E + N_S - n_p N_G \right] + 4(2n_p+1) \Omega_s g_a C_\pm
\\
\dot{N}_E = -2 (\Omega_s g_s + \Omega_a g_a) \cdot [(n_p+1)N_E - n_p N_S] - 4 \Omega_s g_a n_p  C_{\pm} + 2 \gamma_p N_S - ( 2\gamma_1 + \gamma_u) N_E
\\
\begin{aligned}
\dot{N}_S = (\Omega_s g_s + \Omega_a g_a) [n_p N_G + (n_p+1)N_E - (2n_p+1)N_S] - (\Omega_s g_a + \Omega_a g_s) C_\pm +\\+ \gamma_p(N_G-N_S) + \gamma_1 (N_E-N_S) + \frac{\gamma_u}{2} N_E
\end{aligned}
\\
\dot{N}_G = 2 (\Omega_s g_s + \Omega_a g_a) \cdot [ (n_p+1) N_S - n_p N_G ] + 4 g_a \Omega_s (n_p+1) C_\pm - 2 \gamma_p N_G + 2 \gamma_1 N_S
\end{cases}.
\end{equation}
\end{widetext}

\subsection{Ensemble averaging}
Let $N$ be the total number of ion pairs in the cavity. As for isolated
ions, we neglect cavity-mediated coherences between different pairs in
the weak-coupling and strong-dephasing limit. The equations for
$N_{E,G,S}$ then apply to the ensemble-averaged pair populations. The terms $g_{a,s} \Omega_{a,s}$ should also be averaged over the ensemble, whose mean values depend on the correlation length of the crystal field:
\[ 
\begin{split}
\langle 2 g_s \Omega_s \rangle &= \frac{3\xi}{2} \cdot \left( \langle \cos^2 \theta_1 \rangle + \langle \cos^2 \theta_2 \rangle + 2 \langle \cos \theta_1 \cos \theta_2 \rangle \right) \\
\langle 2 g_a \Omega_a \rangle &= \frac{3\xi}{2} \cdot \left( \langle \cos^2 \theta_1 \rangle + \langle \cos^2 \theta_2 \rangle - 2 \langle \cos \theta_1 \cos \theta_2 \rangle \right)  \\
\langle 2g_a \Omega_s \rangle &= \frac{3\xi}{2} \cdot \left( \langle \cos^2 \theta_1 \rangle - \langle \cos^2 \theta_2 \rangle \right) = 0
\end{split},\]

However, in the averaged rate equations \eqref{final-single-pair} for $N_E$ and $N_G$ we have only expressions $\langle 2 g_s \Omega_s \rangle + \langle 2 g_a \Omega_a \rangle = 2 \xi$ and $\langle 2 g_s \Omega_a \rangle=0$ that do not depend on microscopic correlations. Consequently, in the strong-dephasing limit, pair emission becomes equivalent to emission from two independent ions, without Dicke superradiant or subradiant contributions. The correlator $C_{\pm}$  also decouples and does not affect the dynamics.

The equation for the number of photons $n_p$ contains a contribution from each pair. The resulting rate equations for the ensemble of pairs are:
\begin{equation}\label{final-pairs}
\begin{cases}
\dot{n}_p = -\gamma n_p + 2 \xi N  n_p (N_E - N_G) + 2 \xi N  (N_E+N_S)
\\
\dot{N}_E = 2 \xi n_p (N_S - N_E) + 2 \gamma_p N_S - [2(\gamma_1+\xi) + \gamma_u] N_E
\\
\dot{N}_G = 2 \xi  n_p ( N_S -  N_G ) - 2 \gamma_p N_G + 2 (\gamma_1+\xi) N_S
\\
N_S = (1-N_E-N_G)/2
\end{cases}.
\end{equation}
The spontaneous emission corrections in the equations for $N_{E,G}$ can be absorbed into a redefinition of $\gamma_1$ as was possible for isolated ions.

\section{Resulting rate equations}
Isolated ions and ion pairs interact only through the cavity and we can merge the systems \eqref{final-single-ions} and \eqref{final-pairs}. Let $N$ be the total number of ions. These ions form $\chi_pN$ pairs, while $(1-2\chi_p)N$ ions remain isolated. Thus, $\chi_p$ is the number of pairs per ion, $0\leq\chi_p\leq1/2$, and $2\chi_p$ is the fraction of ions belonging to pairs. For the resulting system, we introduce the photon number per ion $n=n_p/N$ and the laser Rabi frequency $\Omega_0 = N \xi$. Also we redefine $\gamma_1 \to \gamma_1 + \xi$ to absorb the spontaneous-emission term into the population equations.
\begin{equation}\label{unified-system}
\begin{cases}
\begin{aligned}
\dot{n} = -\gamma n + (1-2 \chi_p)\Omega_0 n D + 2 \chi_p \Omega_0 n (N_E - N_G) + \\ + \xi \left( (1-2 \chi_p) \frac{1+D}{2} +2  \chi_p (N_E+N_S) \right)
\end{aligned}
\\
\dot{D} = -\gamma_1 (1+D) + \gamma_p (1-D) - 2 \Omega_0 nD
\\
\dot{N}_E = 2 \Omega_0 n (N_S - N_E) + 2 \gamma_p N_S - ( 2\gamma_1 + \gamma_u) N_E
\\
\dot{N}_G = 2 \Omega_0  n( N_S -  N_G ) - 2 \gamma_p N_G + 2 \gamma_1 N_S
\\
N_S = (1 - N_E - N_G)/2
\end{cases}
\end{equation}

To  investigate the system \eqref{unified-system} numerically we establish constants based on experimental papers. Rates $\gamma=10^8~\text{s}^{-1}$ and $\gamma_1=10^2~\text{s}^{-1}$ were discussed above in the context of the rate hierarchy. The range of $\chi_p$ values in which nontrivial EDFL dynamics has been observed is of particular interest: $\chi_p \approx 0.1$ \cite{LeBoudec1993-different-fibers}.

The value of $\Omega_0$ is estimated from the power-gain coefficient as
$\Omega_0=gc/n_g$. For a fiber gain of approximately
$10~\mathrm{dB/m}$ \cite{Smirnov2025} and $n_g\approx1.47$, this gives
$\Omega_0\approx5\times10^8~\mathrm{s}^{-1}$. The corresponding
$\xi=\Omega_0/N\approx10^{-4}~\mathrm{s}^{-1}$ implies
$N\approx5\times10^{12}$ ions in the cavity mode. The values of the
constants are summarized in Table~\ref{tab:parameters}.

\begin{table}[b]
\caption{\label{tab:parameters}Parameters used in the numerical modeling.}
\begin{ruledtabular}
\begin{tabular}{ccccccc}
$\gamma$ & $\gamma_1$ & $\Omega_0$ & $\xi$ & $\gamma_p/\gamma_1$ & $\chi_p$ & $\gamma_u$ \\
\colrule
$10^8~\text{s}^{-1}$ & $10^2~\text{s}^{-1}$ & $5 \cdot 10^{8}~\text{s}^{-1}$ & $10^{-4}~\text{s}^{-1}$ & \multicolumn{3}{c}{Vary} \\
\end{tabular}
\end{ruledtabular}
\end{table}

\begin{figure}[t]
\centering
    \includegraphics[width=.9\columnwidth]{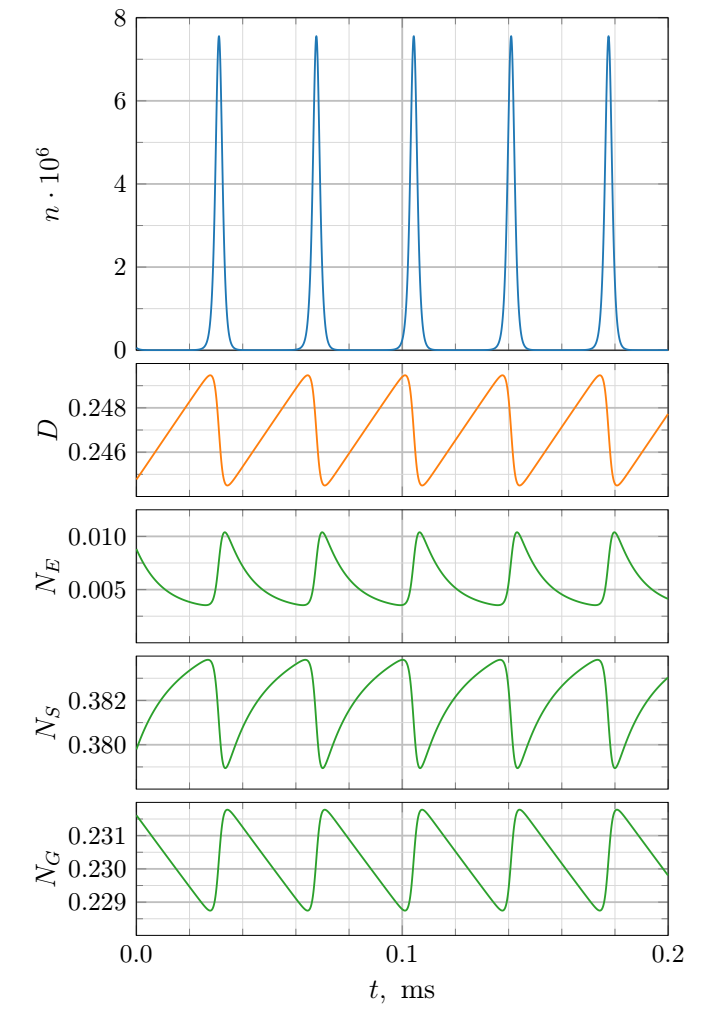}
     \caption{\label{fig:self-pulsing}Self-pulsing regime of the system for $\chi_p=0.05$, $\gamma_u=10^5~\text{s}^{-1}$ and $\gamma_p=4\gamma_1$. Photon number per ion $n$, population inversion $D$, populations of pair states $N_{E,G,S}$.}
\end{figure}

For $\chi_p=0$ the system \eqref{unified-system}  reduces to the class-B laser rate equations with two degrees of freedom allowing only for continuous-wave (CW) radiation with a threshold $\gamma_p \approx \gamma_1$ while $\gamma \ll \Omega_0$. It is well known that additional degrees of freedom can result in the appearance of a complicated phase-space with bistability and limit cycles \cite{Sanchez1996,Otupiri2020}. In our system, pairs serve as this additional degree of freedom that makes a stationary point unstable. Like the Sanchez model  \cite{Sanchez1996}, our system \eqref{unified-system} loses stability of its stationary point and exhibits self-pulsing. Figure~\ref{fig:self-pulsing} shows an example of this regime. The shape of pulses qualitatively coincides with those reported in previously published papers \cite{Sanchez1993,LeBoudec1993,Smirnov2025}.

\subsection{Limit cycle search}
The rates in the system \eqref{unified-system} differ by orders of
magnitude. Consequently, convergence to the periodic regime may require
hundreds of periods, and direct integration is inefficient for locating a
limit cycle. We therefore compute the periodic orbit as a boundary-value
problem. During the cycle, the photon number per ion $n$ varies by several
orders of magnitude, while the other variables vary only slightly. The
substitution $u=\ln n$ regularizes the problem and provides uniform
relative accuracy for the photon number.

First, we have to determine a stationary point $(u^*,D^*,N_E^*, N_G^*)$ of the system and eigenvalues $\lambda_k$ of the Jacobian at that point. For $\chi_p=0$ the stationary point can be found analytically, and we use it as  an initial approximation for a numerical solution of the full problem. In case of poor convergence, we increase $\chi_p$ in several steps.

Let $\lambda_0$ denote the eigenvalue with the largest real part. Within the present numerical classification, $\Re\lambda_0<0$ corresponds to a stable CW stationary state, while $\Re\lambda_0>0$ indicates its instability. In the unstable cases reported here, direct integration confirms the existence of a stable limit cycle; near a Hopf bifurcation, its frequency is approximately $|\Im\lambda_0|/(2\pi)$.

Second, we take a point corresponding to the stationary point but with "turned off" stimulated emission:
$\left( \ln \left[ \frac{\xi}{\gamma} \left[ (1-2 \chi_p) \frac{1+D^*}{2} +2  \chi_p (N_E^*+N_S^*) \right] \right], D^*, N_E^*, N_G^* \right) $ and directly integrate one loop in the $(u, D)$ plane. The duration of this loop is used as the initial guess for the period $T$, and the trajectory itself is used as the initial guess in a boundary-value problem. Two-point boundary conditions are periodic conditions $u(0)=u(T)$, $D(0)=D(T)$, $N_E(0)=N_E(T)$, $N_G(0)=N_G(T)$ and a phase-fixing condition $D(0)=D^*$. 

Third, we check the convergence of the method using direct integration: an inaccurate limit cycle tends to drift. The described method is mostly stable in the region of interest; difficulties arise only near bifurcation points, where $\Re \lambda_0 \approx 0$ and the limit cycle is small.

For all numerical calculations we use Python with NumPy and SciPy: \texttt{solve\_ivp} with the RK45 method (\texttt{rtol=1e-12}, \texttt{atol=1e-12}) and \texttt{solve\_bvp} (\texttt{tol=1e-8}, \texttt{max\_nodes=20000}).

\subsection{Numerical results}

\begin{figure}[b]
\centering
    \includegraphics[width=.95\columnwidth]{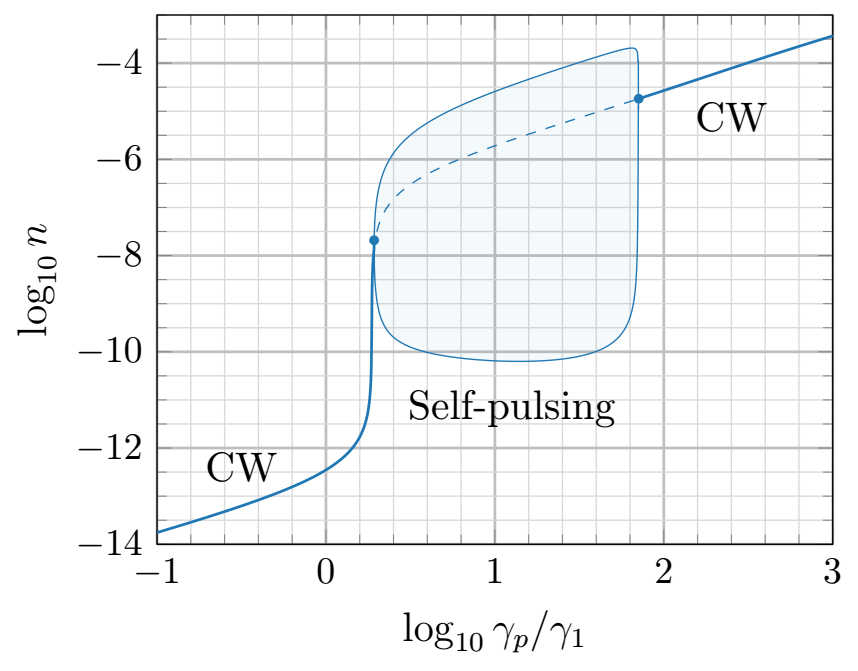}
       \caption{\label{fig:scurve} S-curve for $\gamma_u=10^5~\text{s}^{-1}$, $\chi_p=0.05$ with a self-pulsing region.} 
\end{figure}

\begin{figure}[t]
\centering
    \includegraphics[width=.95\columnwidth]{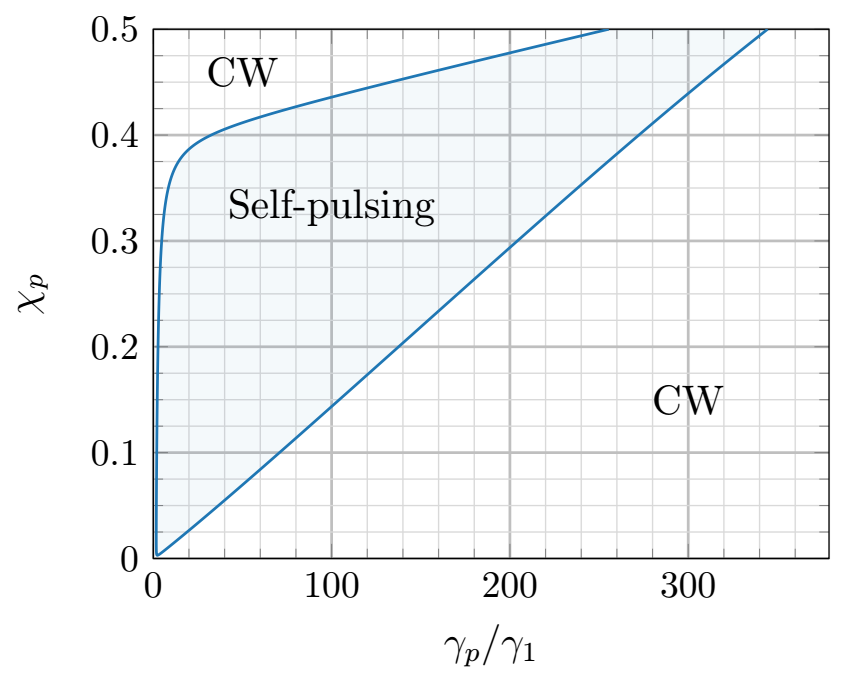}
       \caption{\label{fig:bifurcationD} Bifurcation diagram for $\gamma_u=10^5~\text{s}^{-1}$.}
\end{figure}

In the parameter range studied, the stationary branch of the laser characteristic curve (S-curve) undergoes two Hopf bifurcations, which bound a self-pulsing region. In Fig.~\ref{fig:scurve}, in the CW regime the steady photon number per ion $n$ is plotted; in the self-pulsing regime the region between the maximum $n_\text{max}$  and the minimum $n_\text{min}$  of the limit cycle is plotted. For a fixed value of $\gamma_u$ we obtained a bifurcation diagram (Fig.~\ref{fig:bifurcationD}) showing that the region of pumps hosting self-pulsing grows as $\chi_p$ increases.

\begin{figure}[b]
\centering
    \includegraphics[width=.95\columnwidth]{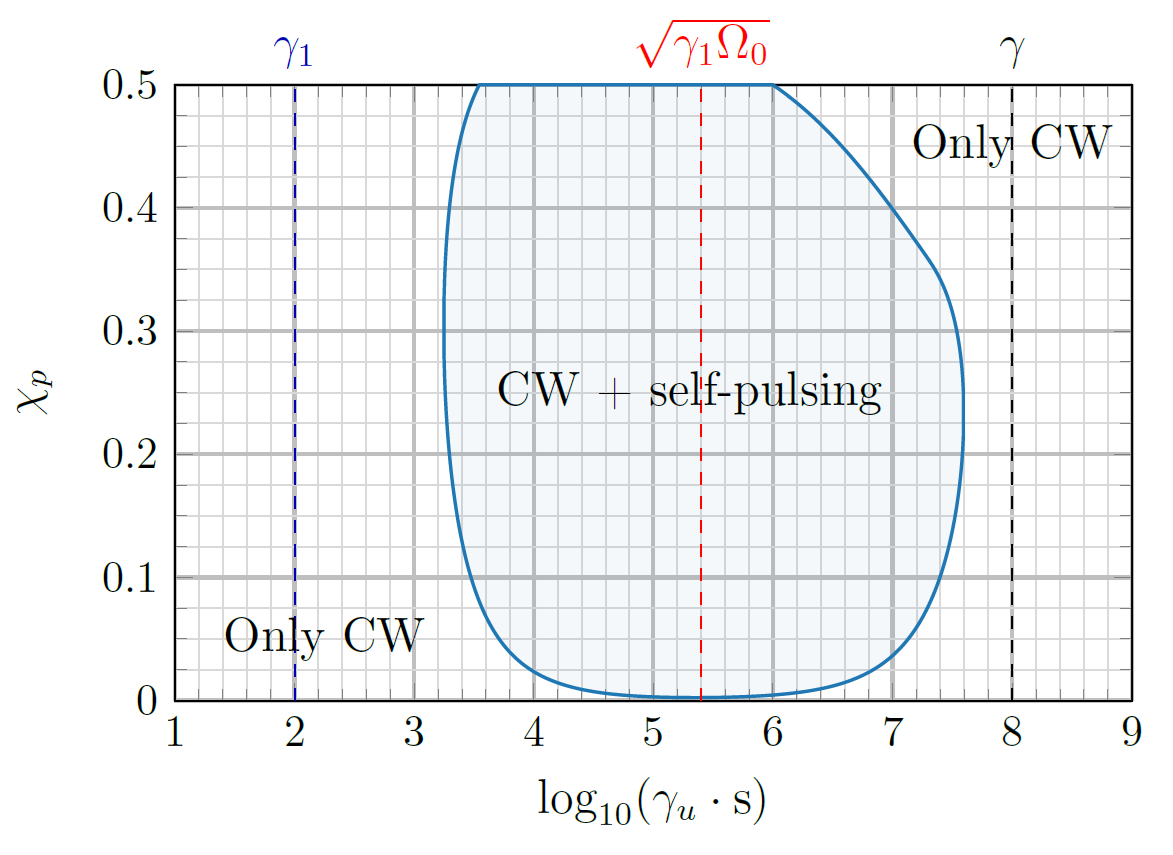}
       \caption{\label{fig:phase-diagram}Phase diagram in the plane $(\gamma_u,\chi_p)$. The inner region marks parameters for which
self-pulsing occurs at some pump rate; outside this region, only CW
operation is found.}
\end{figure}

Among the system parameters, $\gamma_u$ and $\chi_p$ are the least constrained by direct measurements. Therefore, it is important to investigate the region in the plane  $(\gamma_u, \chi_p)$ where the self-pulsing regime is possible in principle for some pump values. We obtained such a diagram (see Fig.~\ref{fig:phase-diagram}), which shows that self-pulsing requires a sufficiently large pair fraction and occurs within a finite interval of $\gamma_u$. In the parameter range studied, this interval lies between $\gamma_1$ and $\gamma$, and $\gamma_u$ is comparable to the angular relaxation-oscillation frequency $\sqrt{\gamma_1\Omega_0\left(\frac{\gamma_p}{\gamma_1}-1\right)}$.

The resulting phase diagram is determined by the rates of the system and can vary for different fibers and cavities. Nevertheless, it is useful to estimate the lower bound on the value of  $\chi_p$ at which self-pulsing appears. Here we reproduce the problem faced by Sanchez et al. \cite{Sanchez1993}: without artificially changing the pair cross-section the model predicts that self-pulsing appears for $\chi_{p,\text{crit}}\approx 10^{-3}$, which is much smaller than the experimental estimate \cite{LeBoudec1993-different-fibers}. Our model reveals the lack of microscopic motivation behind the strongly suppressed pair cross-section. Alternatively, we believe that the discrepancy is due to a broad distribution of $\gamma_u$ over the pairs, which enters fluorescence decay and lasing differently. Therefore, Fig.~\ref{fig:phase-diagram} characterizes the present effective model rather than providing a quantitative prediction for experiment.

\begin{figure}[t]
\centering
    \includegraphics[width=.95\columnwidth]{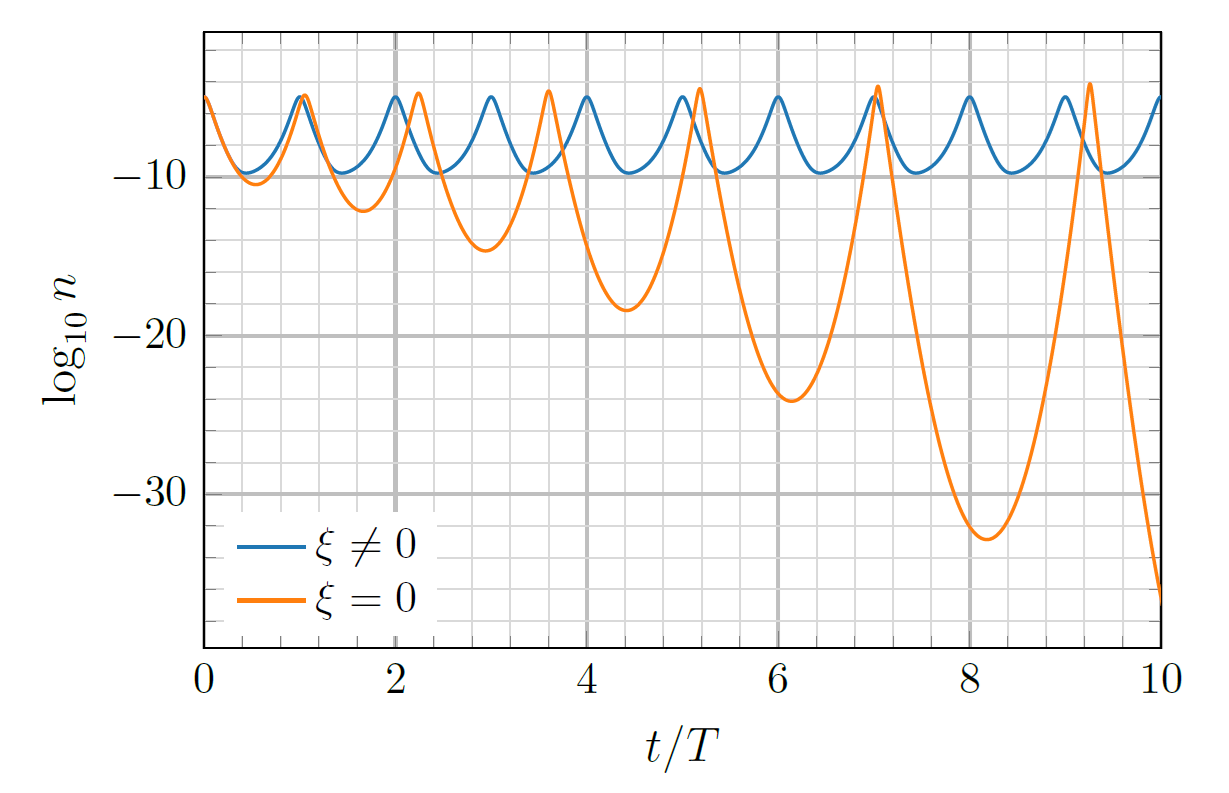}
       \caption{\label{fig:spontaneous-emission}Direct integration of the
system~\eqref{unified-system} with spontaneous emission and with its
source term set to zero at fixed $\Omega_0$.}
\end{figure}

We highlight the importance of the spontaneous-emission term. Although
its rate $\xi$ is orders of magnitude lower than the other rates, it
plays a significant role. Figure~\ref{fig:spontaneous-emission} shows
direct integration of the system~\eqref{unified-system} with and without
the spontaneous-emission source term, starting from the same initial
point. To isolate this contribution, we set $\xi=0$ only in the source
term while keeping $\Omega_0$ fixed. The trajectories initially coincide,
until spontaneous emission starts to act as a ``floor'' that prevents
the photon number from falling to arbitrarily small values between
pulses. Without this source term, the photon-number minima progressively
decrease and the oscillation period increases. Once $n_p=Nn\lesssim1$,
the latter trajectory represents only a formal continuation of the
deterministic mean-field model.

\section{Discussion and Outlook}

In this paper, we have derived a microscopically motivated rate-equation model for a heavily doped erbium fiber laser containing isolated ions and interacting ion pairs. Starting from the Lindblad master equation, we incorporate cooperative upconversion as an effective dissipative process and obtain a closed system of rate equations through adiabatic elimination and ensemble averaging. In the strong-dephasing limit, the ensemble-averaged pair–cavity coupling reduces to the sum of two independent-ion contributions, with no Dicke-type superradiant or subradiant effects in the resulting rate equations.

Let us rewrite the classical Sanchez rate equations \cite{Sanchez1993} using our notation wherever possible. 
\[
\begin{cases}
\dot{n}_p = - \gamma n_p + y \cdot \xi N  n_p (N_E-N_G) \\
\dot{N}_E = y \cdot \xi n_p (2N_S-N_E) + \Lambda - \gamma_{22}N_E \\
\dot{N}_G = y \cdot \xi n_p (2N_S - N_G) - \Lambda + \gamma_{12}N_S \\
N_S = (1-N_E-N_G)/2
\end{cases}
.\]

Here $\Lambda$ is the pump term, and $\gamma_{22}$ and $\gamma_{12}$
are the relaxation coefficients used in the original Sanchez model. 

The microscopic derivation does not leave $y$ as a free fitting parameter:
in the notation of Sanchez \emph{et al.}, the derived net stimulated
pair--field coupling corresponds to $y=2$. Their experimental fit instead
requires $y\simeq0.2$~\cite{Sanchez1993}, an order of magnitude smaller.
Within the present microscopically controlled approximation, this is not a
matter of parameter calibration: inserting $y\simeq0.2$ directly contradicts
the derived pair--cavity coupling and cannot be done while retaining the
microscopic model. Moreover, the discrepancy cannot be removed by redefining
$y$, because the population-transfer and pump terms also differ from the
microscopic equations. The fitted Sanchez model therefore represents a
structurally different phenomenological dynamics, rather than a
parametrization of the present model. Our derivation thus reframes the role
of ion pairs in the EDFL rate equations: their contribution is determined by
the microscopic pair--cavity coupling and cooperative upconversion, not by
an independently adjustable suppression of the pair cross-section.

Numerical analysis of the derived equations reproduces self-pulsing within finite intervals of pump rates and identifies the instability region in the parameter space of the upconversion rate and pair fraction. In the parameter range studied, self-pulsing is found when the upconversion rate is comparable to the characteristic relaxation-oscillation frequency. However, the predicted critical pair fraction is substantially smaller than the experimental estimate. This discrepancy suggests that additional effects, such as a distribution of upconversion rates, inhomogeneous coupling, or more complex ion aggregates, may be important for quantitative agreement with experiments.

Furthermore, our model explicitly includes spontaneous emission. Its
importance as a physically motivated regularization term in classical rate
equations has been emphasized repeatedly~\cite{Okhotnikov1995,Sujecki2018,Smirnov2018}. In the present model, spontaneous emission is also essential quantitatively: as shown in
Fig.~\ref{fig:spontaneous-emission}, neglecting this source term strongly changes both the photon-number minima and the oscillation period. Its
neglect or inconsistent treatment can therefore substantially undermine the quantitative predictive power of the classical rate-equation model.

\bibliography{apssamp}

@PREAMBLE{
 "\providecommand{\noopsort}[1]{}" 
 # "\providecommand{\singleletter}[1]{#1}%" 
}

@INPROCEEDINGS{Georges1992,
   author       = "T. Georges and E. Delevaque and M. Monerie and P. Lamouler and J.-F. Bayon",
   title        = "Pair induced quenching in erbium doped silicate fibers",
   booktitle    = "Optical Amplifiers and Their Applications",
   pages        = "WE4",
   address      = "Santa Fe, NM",
   month        = jun,
   year         = "1992",
}

@ARTICLE{Delevaque1993,
   author       = "E. Delevaque and T. Georges and M. Monerie and P. Lamouler and J.-F. Bayon",
   title        = "Modeling of pair-induced quenching in erbium-doped silicate fibers",
   journal      = "IEEE Photonics Technology Letters",
   volume       = "5",
   number       = "1",
   pages        = "73--75",
   year         = "1993",
   doi          = "10.1109/68.185065",
}

@ARTICLE{LeBoudec1993,
   author       = "P. Le Boudec and M. Le Flohic and P. L. Fran{\c c}ois and F. Sanchez and G. Stephan",
   title        = "Self-pulsing in $\mathrm{Er}^{3+}$-doped fibre laser",
   journal      = "Optical and Quantum Electronics",
   volume       = "25",
   pages        = "359--367",
   year         = "1993",
   doi          = "10.1007/BF00420691",
}

@ARTICLE{Sanchez1993,
   author       = "F. Sanchez and P. Le Boudec and P. L. Fran{\c c}ois and G. Stephan",
   title        = "Effects of ion pairs on the dynamics of erbium-doped fiber lasers",
   journal      = "Physical Review A",
   volume       = "48",
   pages        = "2220--2229",
   year         = "1993",
   doi          = "10.1103/PhysRevA.48.2220",
}

@ARTICLE{Sanchez1996,
   author       = "F. Sanchez and G. Stephan",
   title        = "General analysis of instabilities in erbium-doped fiber lasers",
   journal      = "Physical Review E",
   volume       = "53",
   pages        = "2110--2122",
   year         = "1996",
   doi          = "10.1103/PhysRevE.53.2110",
}

@ARTICLE{Nonlinear1994,
   author       = "P. Xie and S. C. Rand",
   title        = "Nonlinear dynamics of cooperative upconversion",
   journal      = "Journal of the Optical Society of America B",
   volume       = "11",
   number       = "5",
   pages        = "901--912",
   year         = "1994",
   doi          = "10.1364/JOSAB.11.00901",
}

@ARTICLE{Vovchenko2024,
   author       = "I. V. Vovchenko and E. S. Andrianov and A. V. Dorofeenko",
   title        = "Open Quantum System Model for Describing Erbium Laser Operation",
   journal      = "Journal of Radioelectronics",
   number       = "4",
   year         = "2024",
   doi          = "10.30898/1684-1719.2024.4.6",
   note         = "(in Russian)",
}

@INPROCEEDINGS{Vovchenko2025,
   author       = "I. V. Vovchenko and E. S. Andrianov and A. V. Dorofeenko and A. M. Smirnov and O. V. Butov",
   title        = "Full-quantum Description of Erbium-doped Laser with Consideration of Ion Pairs Effects",
   booktitle    = "PIERS 2025, Abu Dhabi",
   month        = may,
   year         = "2025",
}

@ARTICLE{LeBoudec1993-different-fibers,
   author       = "P. Le Boudec and P. L. Fran{\c c}ois and E. Delevaque and J. F. Bayon and F. Sanchez and G. M. Stephan",
   title        = "Influence of ion pairs on the dynamical behaviour of $\mathrm{Er}^{3+}$-doped fibre lasers",
   journal      = "Optical and Quantum Electronics",
   volume       = "25",
   number       = "8",
   pages        = "501--507",
   year         = "1993",
}

@ARTICLE{Snoeks1995,
   author       = "E. Snoeks and G. N. Van den Hoven and A. Polman and B. Hendriksen and M. B. J. Diemeer and F. Priolo",
   title        = "Cooperative upconversion in erbium-implanted soda-lime silicate glass optical waveguides",
   journal      = "Journal of the Optical Society of America B",
   volume       = "12",
   number       = "8",
   pages        = "1468--1474",
   year         = "1995",
   doi          = "10.1364/JOSAB.12.001468",
}

@ARTICLE{Otupiri2020,
   author       = "R. Otupiri and B. Krauskopf and N. G. Broderick",
   title        = "The Yamada Model for a Self-Pulsing Laser: Bifurcation Structure for Nonidentical Decay Times of Gain and Absorber",
   journal      = "International Journal of Bifurcation and Chaos",
   volume       = "30",
   number       = "14",
   pages        = "2030039",
   year         = "2020",
   doi          = "10.1142/S0218127420300399",
}

@ARTICLE{Smirnov2025,
   author       = "A. M. Smirnov and A. A. Rybaltovsky and I. A. Nechepurenko and A. V. Dorofeenko and O. V. Butov",
   title        = "Critical behavior of erbium fiber lasers in pulsed and relaxation oscillations regimes",
   journal      = "Optics \& Laser Technology",
   volume       = "181",
   pages        = "111823",
   year         = "2025",
   doi          = "10.1016/j.optlastec.2024.111823",
}

@BOOK{Scully,
   author       = "M. O. Scully and M. S. Zubairy",
   title        = "Quantum Optics",
   publisher    = "Cambridge University Press",
   year         = "1997",
}

@ARTICLE{Okhotnikov1995,
   author       = "O. G. Okhotnikov and J. R. Salcedo",
   title        = "Emission buildup in Q-switched fiber lasers",
   journal      = "Optics Letters",
   volume       = "20",
   number       = "8",
   pages        = "887--889",
   year         = "1995",
   doi          = "10.1364/OL.20.000887",
}

@ARTICLE{Sujecki2018,
   author       = "S. Sujecki",
   title        = "Numerical Analysis of Q-Switched Erbium Ion Doped Fluoride Glass Fiber Laser Operation Including Spontaneous Emission",
   journal      = "Applied Sciences",
   volume       = "8",
   number       = "5",
   pages        = "803",
   year         = "2018",
   doi          = "10.3390/app8050803",
}

@ARTICLE{Smirnov2018,
   author       = "A. M. Smirnov and A. P. Bazakutsa and Y. K. Chamorovskiy and I. A. Nechepurenko and A. V. Dorofeenko and O. V. Butov",
   title        = "Thermal Switching of Lasing Regimes in Heavily Doped $\mathrm{Er}^{3+}$ Fiber Lasers",
   journal      = "ACS Photonics",
   volume       = "5",
   number       = "12",
   pages        = "5038--5046",
   year         = "2018",
   doi          = "10.1021/acsphotonics.8b01298",
}

@ARTICLE{Lacot1994,
   author       = "E. Lacot and F. Stoeckel and M. Chenevier",
   title        = "Dynamics of an erbium-doped fiber laser",
   journal      = "Physical Review A",
   volume       = "49",
   number       = "5",
   pages        = "3997--4008",
   year         = "1994",
   doi          = "10.1103/PhysRevA.49.3997",
}

@ARTICLE{Babin2007,
   author       = "S. A. Babin and D. V. Churkin and A. E. Ismagulov and S. I. Kablukov and M. A. Nikulin",
   title        = "Single frequency single polarization DFB fiber laser",
   journal      = "Laser Physics Letters",
   volume       = "4",
   number       = "6",
   pages        = "428--432",
   year         = "2007",
}

@ARTICLE{Yamashita1998,
   author       = "S. Yamashita and K. Hsu",
   title        = "Single-frequency, single-polarization operation of tunable miniature erbium:ytterbium fiber {F}abry--{P}\'erot lasers by use of self-injection locking",
   journal      = "Optics Letters",
   volume       = "23",
   number       = "15",
   pages        = "1200--1202",
   year         = "1998",
   doi          = "10.1364/OL.23.001200",
}

@book{breuer2002theory,

title={The theory of open quantum systems},

author={Breuer, Heinz-Peter and Petruccione, Francesco},

year={2002},

publisher={OUP Oxford}

}

@article{Layne1977,
  title = {Multiphonon relaxation of rare-earth ions in oxide glasses},
  author = {Layne, C. B. and Lowdermilk, W. H. and Weber, M. J.},
  journal = {Phys. Rev. B},
  volume = {16},
  issue = {1},
  pages = {10--20},
  numpages = {0},
  year = {1977},
  month = {Jul},
  publisher = {American Physical Society},
  doi = {10.1103/PhysRevB.16.10},
  url = {https://link.aps.org/doi/10.1103/PhysRevB.16.10}
}

@book{Siegman1986,
  title={Lasers},
  author={Siegman, Anthony E},
  year={1986},
  publisher={University science books}
}

\end{document}